\documentclass[seceq]{ptptex}

\usepackage{graphicx}
\usepackage{wrapft}

\title{
Testing general relativity with the multipole spectra 
of the SDSS luminous red galaxies
}

\author{
\textsc{Kazuhiro Yamamoto}${}^1$, \textsc{Takahiro Sato}${}^1$, 
\textsc{Gert H\"utsi}${}^{2,3}$}

\inst{
$^1$Department of Physical Sciences, Hiroshima University, 
Higashi-hiroshima, 739-8526, Japan
\\
$^{2}$Department of Physics and Astronomy, University College London, London, WC1E 6BT\\
$^{3}$Tartu Observatory, EE-61602 T\~oravere, Estonia
}

\abst{
As a test of general relativity on cosmological scales, we 
measure the $\gamma$ parameter for the growth rate of density 
perturbations using the redshift-space distortion of the luminous 
red galaxies in the Sloan Digital Sky Survey (SDSS). 
Assuming the cosmological constant model, which 
matches the results of the WMAP experiment, 
{ we find  $\gamma=0.62+1.8(\sigma_8-0.8) \pm 0.11$ at 1-sigma 
confidence level, which is consistent with the prediction of 
general relativity, $\gamma\simeq0.55\sim0.56$.
Rather high value of $\sigma_8(\geq0.87)$ is required 
to be consistent with the prediction of the cosmological 
DGP model, 
$\gamma\simeq0.68$}.}

\begin{document}

\maketitle

{\it Introduction}~--~
Modified gravity models, e.g., $f(R)$ gravity, TeVeS theory, 
DGP model, have been proposed as possible alternatives 
to the dark energy model. Although these models might not be complete, 
they pose ambitious challenges to the fundamental physics, 
suggesting one to go beyond the standard model. 
In fact, a lot of dark energy surveys are already under 
progress.\cite{DETF,European}
Testing the theory of gravity on cosmological scales will surely 
become one of the important objectives of these future large surveys. 
\cite{Linder2006,IshakSpergel,YamamotoA,YamamotoB}.

Measurement of the growth of density perturbations 
will be the key for testing the gravity theory. 
\cite{Linder2005,Amendola,Heavens,MaartensKoyama,KunzSapone,
JainZhang,Chiba,Koyama08,Uzan}. Several authors have already investigated 
the growth of density perturbations as a way of
constraining these theories.
\cite{NesserisPerivolaropoulos,PortoAmendola,Wang}.
In the future weak lensing statistics will be a promising probe of the
density perturbations, while the redshift-space distortions 
may also be useful for constraining the growth rate of 
perturbations \cite{Linder,SaponeAmendola}.
Recently, Guzzo, et al. have reported a constraint on the growth rate
by evaluating the anisotropic correlation function of the galaxy 
sample from the VIMOS-VLT Deep Survey (VVDS) \cite{Guzzo}.

The characteristic redshift of the VVDS galaxy sample is rather large. 
However, the survey area of the VVDS sample is small.
This is a disadvantage in detecting the linear redshift-space
distortions. In the present paper, we use the Sloan Digital Sky
Survey (SDSS) luminous red galaxy (LRG) sample from the Data Release 5, 
whose survey area is around 5000 square degrees. 
In this letter, we present the results of the multipole power 
spectrum analysis for the SDSS LRG sample, and subsequently use it to
measure the $\gamma$ parameter for the growth rate of density 
perturbations. This gives a simple test of general relativity. 
Throughout this paper, we use the units where the light speed 
equals 1.

{\it Measurement of the quadrupole spectrum}~--~
The (linear) growth rate is defined by $f=d\ln D_1/d\ln a$, where 
$D_1$ is the growth factor, and $a$ is the scale factor. 
Due to the continuity equation of the matter density the linear
velocity field is related to the time derivative of the 
matter density contrast, which itself is proportional to the growth 
factor $D_1$.
The peculiar velocity of galaxies contaminates the observed 
redshift, which leads to the difference in the radial position
if the redshift is taken as the indicator of the distance. 
This causes the difference in the spatial clustering 
between redshift space and real space, which is called
the redshift-space distortion. The power spectrum including
the redshift-space distortion can be modeled as (e.g., \citen{PD96PD94})
$
P(k,\mu)=(b(k)+f\mu^2)^2 P_{\rm mass}(k) {\cal D}(k,\mu)$,
where $\mu$ is the directional cosine between the line of sight
direction and the wave number vector, $b(k)$ is the bias factor, 
$P_{\rm mass}(k)$ is the mass power spectrum, ${\cal D}(k,\mu)$ 
describes the damping factor due to the finger of God effect. 

Thus, the redshift-space distortion causes the anisotropy
of the clustering amplitude depending on $\mu$. 
The multipole power spectra are defined by the coefficients 
of the multipole expansion \cite{TH,YamamotoN}
$P(k,\mu)=\sum_{\ell=0,2,\cdots} P_\ell(k) {\cal L}_\ell(\mu)(2\ell+1)$,
where ${\cal L}_\ell(\mu)$  are the Legendre 
polynomials.\footnote{Note that our definition of the 
multipole spectrum $P_\ell(k)$ is different from the conventional 
definition by the factor $2\ell+1$.} 
The monopole $P_0(k)$ represents the angular averaged
power spectrum and is usually what we mean by the power 
spectrum. $P_2(k)$ is the quadrupole  spectrum, which gives
the leading anisotropic contribution. The usefulness of
the quadrupole spectrum for the dark energy is discussed in 
\citen{YamamotoPRL}.

Within the linear theory of density perturbations, 
the ratio of the quadrupole to the monopole is given by
${P_2(k)/ P_0(k)}=
[{4\beta/ 3}+{4\beta^2/7}]/
5[1+{2\beta/3}+{\beta^2/ 5}]$,
where $\beta=f/b(k)$.
Thus, we can measure the growth rate from the quadrupole-monopole ratio. 
However, there are two difficulties to perform this method. 
The first is the clustering bias $b(k)$, for which we need other 
independent information. 
The second is the effect of nonlinear velocity, the finger 
of God effect, which significantly contaminates the quadrupole spectrum,
as one can see in Figure 1-b. 

In the present work we measured the monopole and quadrupole 
power spectra in the clustering of the SDSS DR5 luminous red 
galaxy sample. The galaxy sample used in our analysis
consists of 64867 galaxies over the survey area of 4780 deg${}^2$
and redshift range $0.16\leq z\leq 0.47$.\cite{HutsiAB}. 
We have excluded the southern survey stripes since these just increase the 
sidelobes of the survey window without adding much of the extra volume. 
We have also removed some minor parts of the LRG sample to obtain more 
continuous and smooth chunk of volume.

The scheme to measure the monopole and the quadrupole
spectrum is the same as the one described in reference \citen{YamamotoN}. 
In this analysis we adopted the comoving distance of the 
fiducial model, which we assumed to be a flat universe with the 
cosmological constant and matter density parameter $\Omega_m=0.3$. 
In our power spectrum analysis we fixed 
the weight function equal to 1. 

Figure 1-a shows the monopole spectrum. The error 
bars correspond to the $1$-sigma errors. The result is consistent 
with the previous measurement by H\"utsi in reference \citen{HutsiAB}, 
which is also plotted in this figure for comparison. 
As discussed in reference \citen{HutsiAB}, the monopole spectrum 
reveals the baryon acoustic oscillation feature. 

Figure 1-b plots the ratio of the quadrupole to the monopole.
The quadrupole spectrum changes the signature at the wave number
$k\sim 0.3~h{\rm Mpc}^{-1}$. This is because the finger of God 
effect becomes significant on small scales. 
The solid curve is the theoretical curve for the 
$\Lambda$CDM model with $\Omega_m=0.3$,
$n_s=0.96$ (initial spectral index),  
$\sigma_8=0.8$, $h=0.7$, $\gamma=0.56$ and $\sigma_v=360$km/s
(see also below), and assumes a clustering bias 
$b(k)=1.2[1+0.2(k/0.1h{\rm Mpc}^{-1})^{1/2}]$.

\begin{figure}
  \leavevmode
  \begin{center}
    \begin{tabular}{ c c }
      \includegraphics[width=2.6in,angle=0]{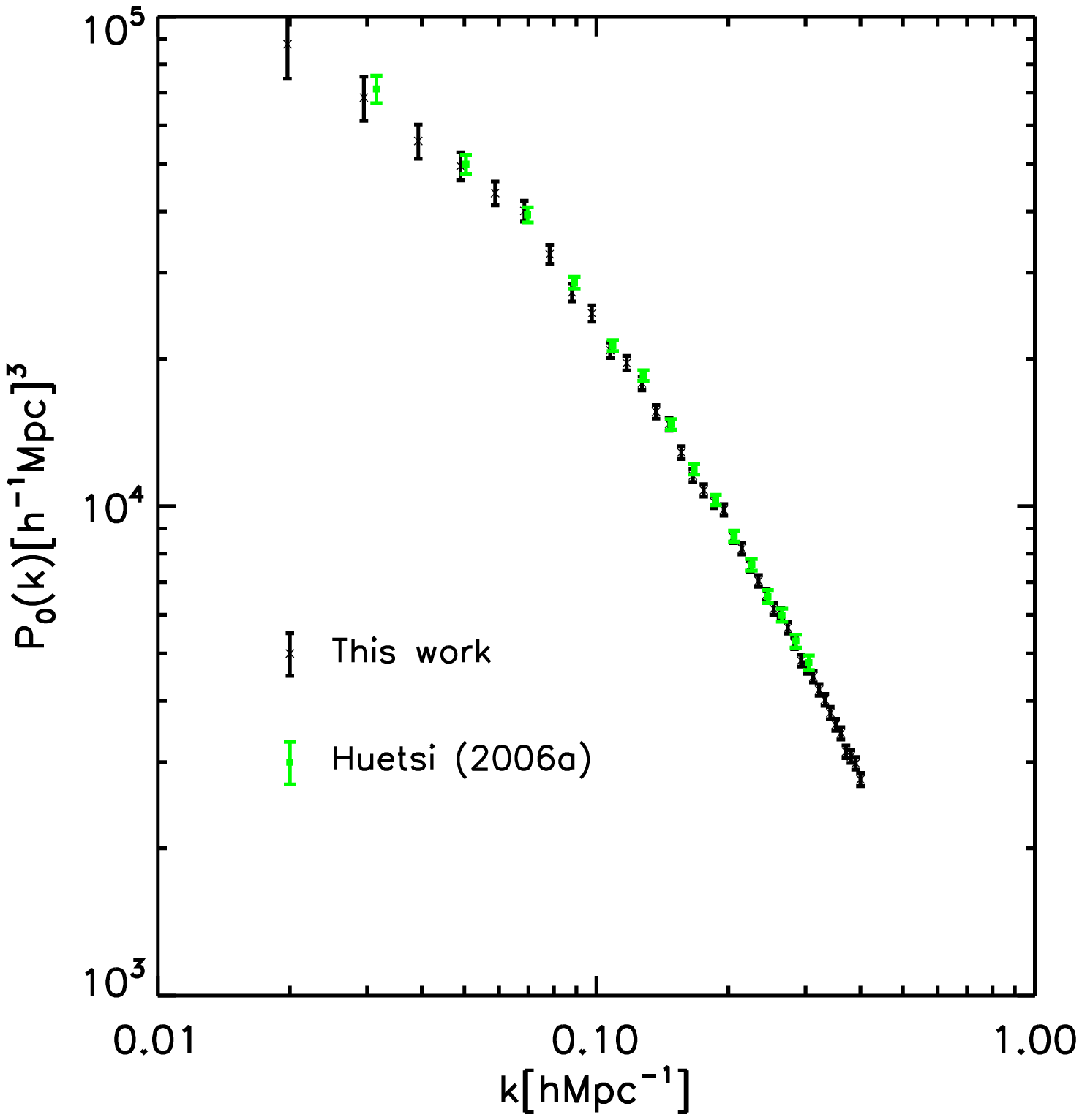}
      &
      \includegraphics[width=2.6in,angle=0]{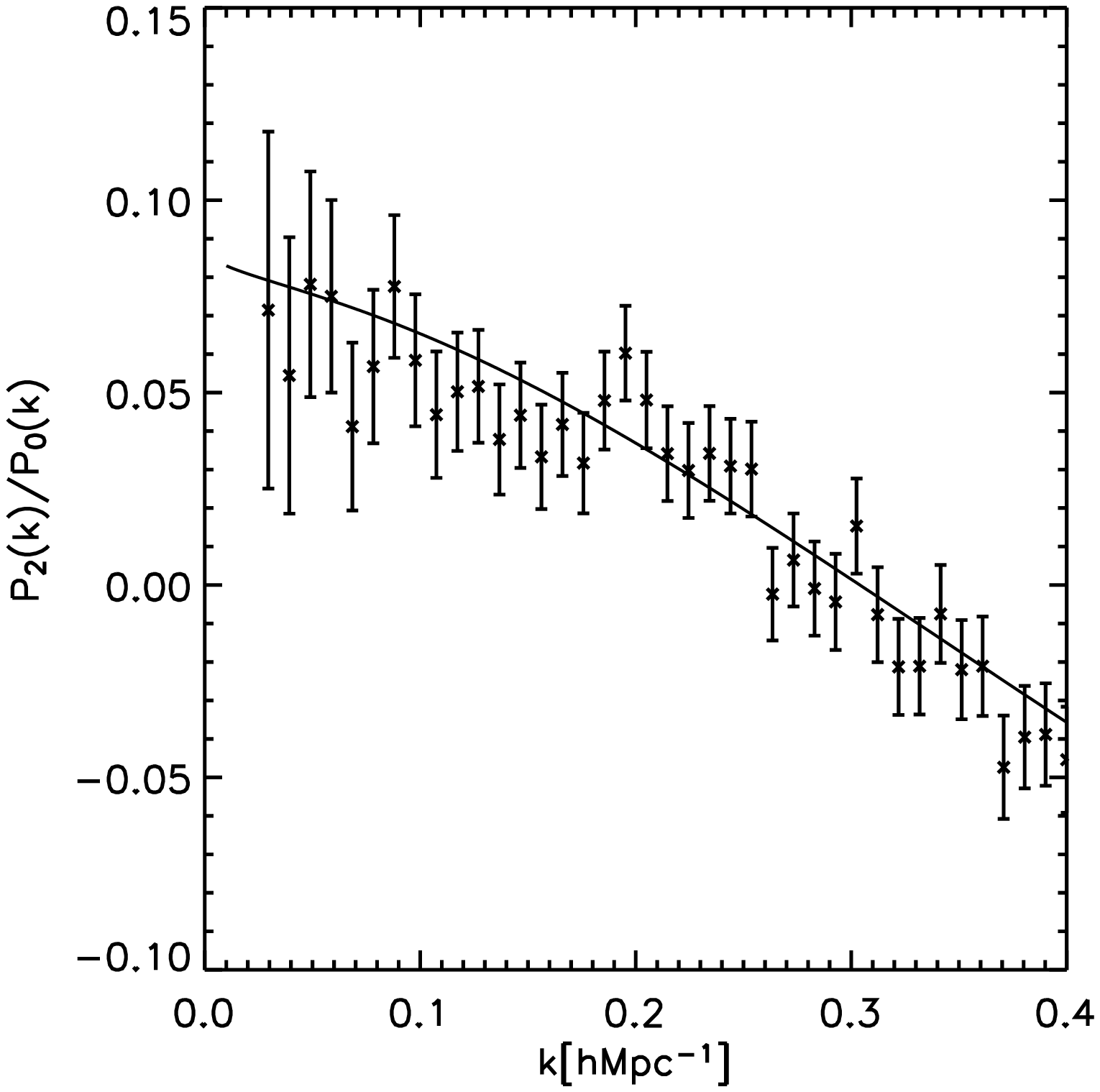}
    \end{tabular}
\caption{ (a,~Left) 
Monopole power spectrum as a function of the wave number. 
The dark (black) points represent this work,  the light (green) points 
the previous result \cite{HutsiAB}. (b,~Right) Quadrupole power spectrum 
divided by the monopole spectrum. The solid curve is the theoretical 
curve of the $\Lambda$CDM model.}
\label{fig:modifiedgravityone}
\end{center}
\end{figure}

{\it Testing the growth rate}~--~
By using the quadrupole spectrum, we perform a simple test
of the gravity theory. In particular, we focus on the $\gamma$ parameter, 
which is introduced to parameterise the growth rate as, 
$  f\equiv{d\ln D_1(a)/ d\ln a} = \Omega_m(a)^\gamma$,
where $\Omega_m(a)=H_0^2\Omega_m a^{-3}/H(a)^2$, $H(a)=\dot a/a$ is 
the Hubble expansion rate, $H_0(=100 h{\rm km/s/Mpc})$ 
is the Hubble parameter.

Measurement of $\gamma$ provides a simple test of the gravity theory. 
Within general relativity, even with the dark energy component, 
$\gamma$ takes the value around $\gamma\simeq0.55$ \cite{Linder2005}. 
However, $\gamma$ may take different
values in modified gravity models. For example, $\gamma\simeq 0.68$, 
in the cosmological DGP model including a self-acceleration 
mechanism. Thus, the measurement of $\gamma$ is a simple
test of general relativity. 

As mentioned in the previous section, we need to take the 
clustering bias and the finger of God effect into account. 
For the finger of God effect we adopt the following 
form of ${\cal D}(k,\mu)$, the damping due to the nonlinear 
random velocity,
 ${\cal D}(k,\mu)={1/[ 1+(k\mu \sigma_v/H_0)^2/2]}$,
where $\sigma_v$ is the one dimensional pairwise velocity dispersion
(e.g., \citen{MJB}). 
This form of the damping assumes an exponential distribution function 
for the pairwise peculiar velocity. 
In order to determine the clustering bias, we simply 
fix the value of $\sigma_8$. If  $\sigma_8$ is fixed, and 
the cosmological parameters and the clustering bias are given, 
we can compute the monopole spectrum $P_0^{\rm theor}(k)$ with
$P_\ell(k)={1\over 2}\int_{-1}^1 d\mu P(k,\mu) {\cal L}_\ell(\mu)$,
where we use the Peacock and Dodds formula for 
the mass power spectrum $P_{\rm mass}(k)$ \cite{PD96PD94}. 
We determine the clustering bias $b(k_i)$ through the condition 
$P_0^{\rm obs}(k_i)=P_0^{\rm theor}(k_i)$ using a numerical method.
Here $P^{\rm obs}_0(k_i)$ is the measured value of the monopole 
at wave number $k_i$, and $P^{\rm theor}_0(k_i)$ is the 
corresponding theoretical value.
Figure 2 exemplifies the bias obtained by the above method
for the cases of $\sigma_8$ fixed as $\sigma_8=0.9,~0.8,~0.7$.

\begin{wrapfigure}{l}{8.cm}
     \includegraphics[width=2.6in,angle=0]{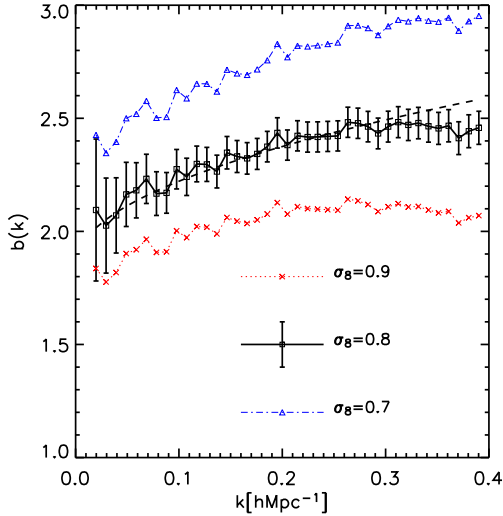}
\caption{
The bias $b(k_i)$ obtained from our numerical method. 
The three curves correspond to $\sigma_8=0.9,~0.8,~0.7$ and 
the other parameters are the same as those of Figure 1-b.
For the case $\sigma_8=0.8$ the errors are estimated 
from those of $P(k_i)$. The dashed curve is 
$b(k)=1.2[1+0.2(k/0.1h{\rm Mpc}^{-1})^{1/2}]$.
}
\label{fig:bias}
\end{wrapfigure}

We used the monopole spectrum to determine the bias, and 
the quadrupole spectrum to obtain constraints on 
$\gamma$ and $\sigma_v$. Since the galaxy sample covers rather
broad redshift range, $0.16\leq z\leq 0.47$, 
the effect of the time-evolution (light-cone effect) should
be considered properly.\cite{YamamotoSuto} 
However, for simplicity, we here evaluated 
the theoretical spectra at the mean redshift of $z=0.31$. 


Figure 3-a demonstrates the contours of $\Delta \chi^2$ in the 
$\gamma$ versus $\sigma_v$ parameter plane. We compute 
$\chi^2$ as 
$\chi^2={\sum}_i{{[P^{\rm obs}_2(k_i)-P^{\rm theor}_2(k_i)]^2}
/{[\Delta P_2^{\rm obs}(k_i)]^2}}$,
where $P^{\rm obs}_2(k_i)$ and $\Delta P^{\rm obs}_2(k_i)$  are the 
measured values and errors as plotted in Figure 1-b. 
$P^{\rm theor}_2(k_i)$ are the corresponding theoretical values. 
The solid curves assume $\sigma_8=0.8$, while the dotted  (dashed)
ones $\sigma_8=0.7$ ($\sigma_8=0.9$). The other 
parameters are fixed as $\Omega_m=0.28$, \footnote{As 
the observed power spectra are obtained by adopting the 
distance-redshift relation of the fiducial model, the flat 
$\Lambda$CDM model with $\Omega_m=0.3$,  
the cosmological distortion effect is taken into account 
properly in our theoretical computation \cite{Ballinger,Matsubara}. } 
$n_s=0.96$, $h=0.7$.\cite{Komatsu}
In Figure 3-a we plot the contour levels of $\Delta\chi^2=2.3$
(inner curves) and 6.2 (outer curves), which correspond to 
the 1-sigma and 2-sigma confidence levels of the $\chi^2$ distribution. 
Clearly, the higher value of $\sigma_8$ favours higher value for $\gamma$. 
{We find  $\gamma=0.62+1.8(\sigma_8-0.8) \pm 0.11~(\pm0.19)$ and 
$\sigma_v=367+80(\sigma_8-0.8) \pm 16~(\pm27)$~km/s 
at $68~(90)$ percent confidence level, respectively.}
The value of $\gamma$ is consistent with general relativity.
The result is not sensitive to the inclusion of the baryon oscillation 
in the theoretical power spectrum. 

The relation of $\gamma$ and $\sigma_8$ can be understood as
the degeneracy between $\sigma_8$ and the growth rate $f$ 
in the following way. As the observed power spectra can be roughly written as 
$P^{\rm obs}_0\propto b^2(k)\sigma_8^2 D^2_1(z)/D^2_1(z=0)$ and 
$P^{\rm obs}_2\propto b(k)f\sigma_8^2 D_1^2(z)/D_1^2(z=0)$, 
the degeneracy between $\sigma_8$ and the growth rate
$f$ (or $\gamma$) in our method is given by
$  f\sigma_8 {D_1(z)/ D_1(z=0)}={\rm constant}$. 

Figure 3-b is the analogue of Figure 3-a, with the expansion 
history now taken to be that of the spatially flat DGP model, 
which follows $H^2(a) -H(a)/r_c=8\pi G\rho/3$, where $\rho$ 
is the matter density and $r_c=H_0/(1-\Omega_m)$ is the 
crossover scale related to the 5-dimensional Planck mass. 
The expansion history in this model can be well approximated by the dark 
energy model with the equation of state parameter $w(a)=w_0+w_a(1-a)$,
where $w_0=-0.78$ and $w_a=0.32$, as long as $\Omega_m\sim 0.3$ 
\cite{Linder2005}. 
However, the Poisson equation is modified, and the growth history 
is approximated by the formula with $\gamma\simeq0.68$.
In order to be consistent with  $\gamma=0.68$, Figure 3-b requires 
higher value of $\sigma_8$ as compared to the $\Lambda$CDM case.
{We find $\gamma=0.47+1.7(\sigma_8-0.8) \pm 0.09$ at $68$ 
percent confidence level, which requires $\sigma_8\geq0.87$}.

The pair-wise velocity dispersion measured in this work is
somewhat smaller than the theoretical model in the reference 
\cite{MJB}.
Li, et al. investigated the pair-wise velocity of the SDSS 
galaxies \cite{Li}. Their analysis is limited since it 
is based on galaxies with redshifts less than 0.3, however,
they report the dependence of the pairwise velocity dispersion on galaxy 
properties and also on scales.

{\it Conclusion}~--~
In summary, we measured the monopole and quadrupole spectra in the spatial 
clustering of the SDSS LRG galaxy sample from DR5. The monopole spectrum 
is consistent with the previous result by H\"utsi. Using the quadrupole
spectrum, we measured the $\gamma$ parameter for the linear growth 
rate and the pair-wise peculiar velocity dispersion. 
The measurement of $\gamma$ provides a simple test of general relativity. 
The measured value of $\gamma$ is consistent with general 
relativity, however, it is inconsistent with the cosmological DGP model, 
$\gamma\simeq0.68$, {as long as $\sigma_8<0.87$}. 
If a constraint on $\sigma_8$ from other independent sources, 
e.g., the cosmic microwave background anisotropies, is included, 
we would be able to obtain tighter constraint on the DGP model 
\cite{Starkman}.
The constraint on $\gamma$ can be applied to other modified 
gravity models, given that the value of $\gamma$ which 
characterizes a particular model is found, as discussed 
by Linder and Cahn \cite{LC} (cf. \cite{PolarskiAPolarskiB}).
In this work we only considered a spatially flat model 
and fixed the cosmological parameters so as to match with 
the results of the WMAP experiment. 
The constraint on $\gamma$ will be weakened by including the 
uncertainties of the cosmological model, e.g., the dark energy 
properties.

\begin{figure}[t]
  \leavevmode
  \begin{center}
    \begin{tabular}{ c c }
      \includegraphics[width=2.4in,angle=0]{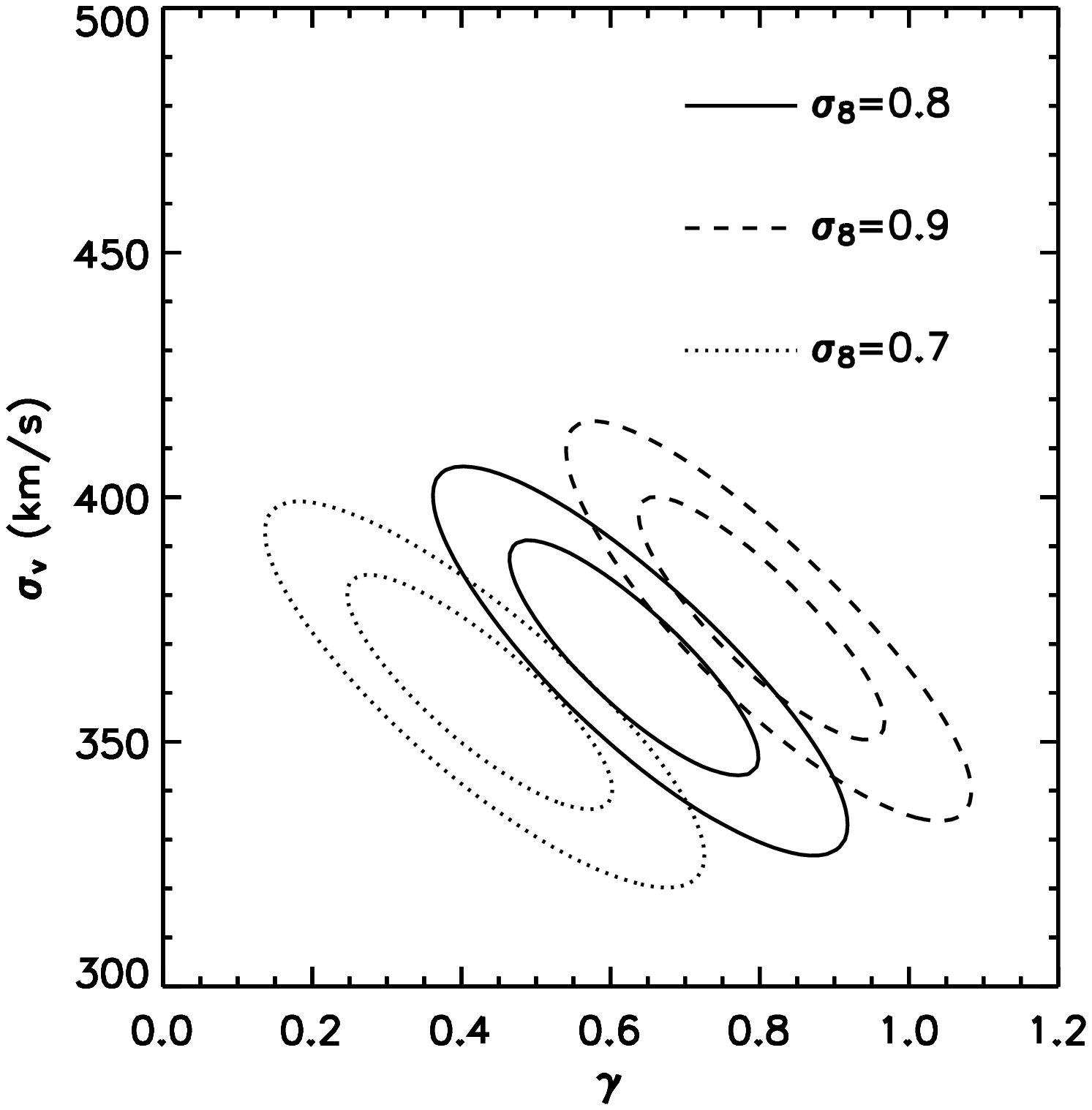}
      &
      \includegraphics[width=2.4in,angle=0]{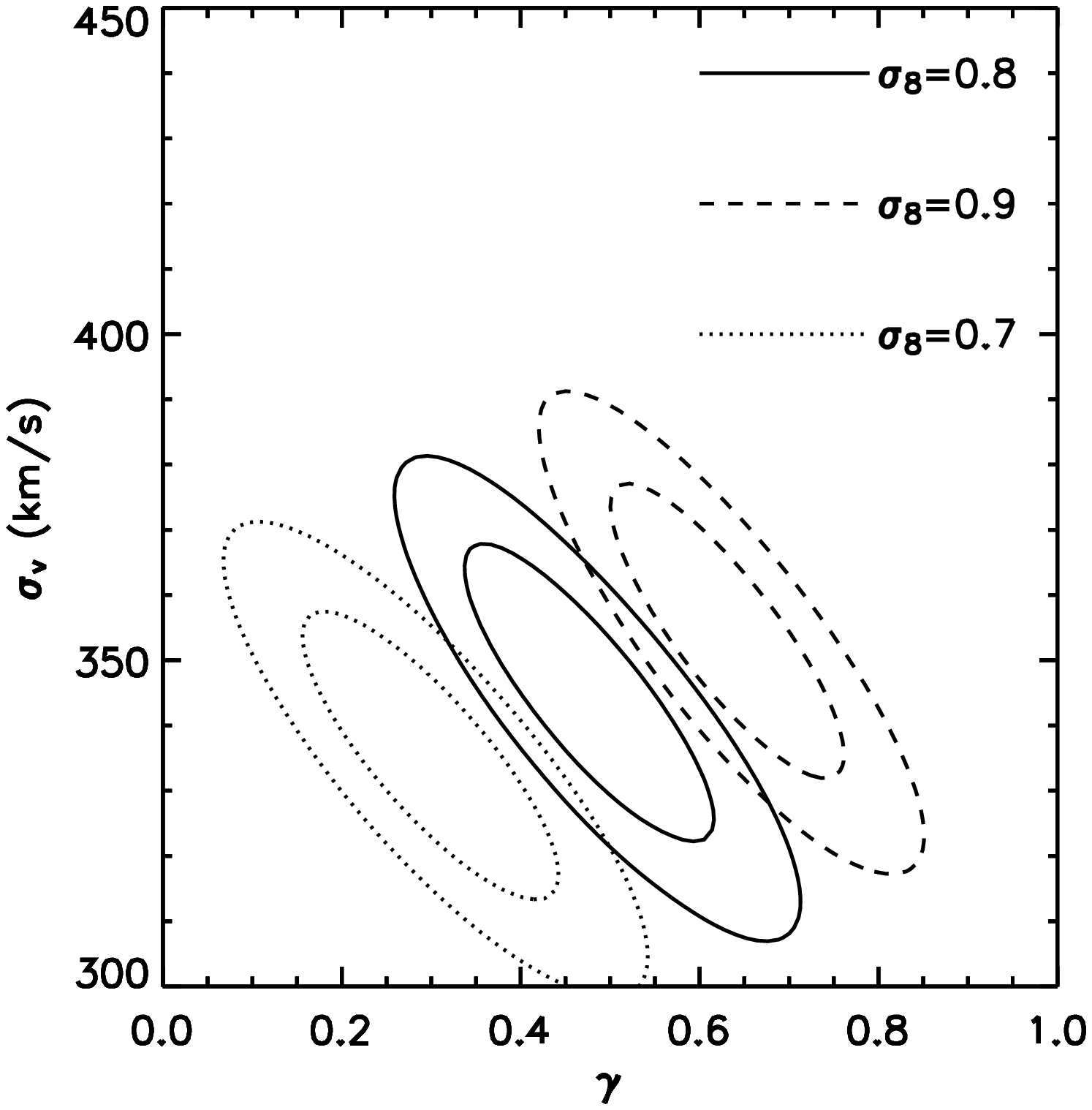}
    \end{tabular}
\caption{ (a,~Left) 
$\Delta\chi^2$ in the $\gamma$-$\sigma_v$ plane. 
We fixed the normalization of the mass power spectrum 
as $\sigma_8=0.7$ (dotted curves), $\sigma_8=0.8$
(solid curves), and $\sigma_8=0.9$ (dashed curves), respectively.
The contour levels are  $\Delta\chi^2=2.3$
(inner curves) and 6.2 (outer curves), which correspond to 
the 1-sigma and 2-sigma confidence levels of the $\chi^2$ distribution. 
The other parameters are fixed as $\Omega_m=0.28$, $n_s=0.96$, and 
$h=0.7$.\cite{Komatsu}
(b,~Right)
Same as the left (a), except here we used the expansion history of 
the DGP model.
}
\end{center}
\end{figure}

\vspace{2mm}
{\it Acknowledgements}~
We would like to thank H.~Nishioka, Y.~Kojima and 
Y.~Suto for useful 
discussions and comments. We thank G.~Nakamura, K.~Koyama and R.~C.~Nichol for 
useful comments and discussions on the earlier version of the manuscript.
This work was supported by a Grant-in-Aid 
for Scientific research of Japanese Ministry of Education, 
Culture, Sports, Science and Technology (No. 18540277).

\end{document}